\documentclass[pras,notitlepage,twocolumn]{revtex4-1}

\usepackage{amsmath}
\usepackage{amsfonts}
\usepackage{graphicx}
\usepackage{subfigure}
\usepackage{pdfsync}
\usepackage{bm}
\usepackage{color}
\usepackage{subfigure}
\usepackage{algorithm}
\usepackage{algpseudocode}
\usepackage{verbatim}
\usepackage{hyperref}

\newcommand{\ket}[1]{\ensuremath {|\: #1 \: \rangle}}
\newcommand{\bra}[1]{\ensuremath{\langle \: #1 \:|}}

\newcommand{\ves}[2]{\ensuremath{#1_1,#1_2, \ldots, #1_{#2}}}

\newcommand{\eref}[1]{(\ref{#1})}
\newcommand{\fref}[1]{figure \ref{#1}}
\newcommand{\Fref}[1]{Figure \ref{#1}}

\newcommand{\llrr}[1]{\ensuremath{\left( #1\right)}}

\begin{document}

\title{A quantum-walk-inspired adiabatic algorithm for graph isomorphism}


\author{Dario Tamascelli}

\affiliation{Dipartimento di Informatica, Universit\`a degli Studi di Milano\\
Via Comelico, 39/41, 20135 Milano- Italy}
\email{tamascelli@di.unimi.it}
\author{Luca Zanetti}
\affiliation{Cluster of Excellence on Multimodal Computing and Interaction, Saarland University, 66123 Saarbr\"ucken, Germany}
\email{luca.zanetti@mpi-inf.mpg.de}



\begin{abstract}
We present a $2$-local quantum algorithm for graph isomorphism GI based on an adiabatic protocol.  By exploiting continuous-time quantum-walks, we are able to avoid a mere diffusion over all possible configurations and to significantly reduce the dimensionality of the visited space.  Within this restricted space, the graph isomorphism problem can be translated into the search of a  satisfying assignment to a $2$-SAT formula without resorting to perturbation gadgets or projective techniques. We present an analysis of the execution time of the algorithm on small instances of the graph isomorphism problem and discuss the issue of an implementation of the proposed adiabatic scheme on current quantum computing hardware. 
\end{abstract}


\maketitle

\section{Introduction}
The graph isomorphism problem (GI) requires to decide whether two given graphs $G_1=(V,E_1)$ and $G_2=(V,E_2)$ are indeed the same graph but for a relabeling of the vertices. Due to its practical applications (ranging from chemistry to social sciences) and theoretical properties, the problem has been thoroughly studied \cite{kst:gip}. 
GI possesses peculiar features that make it an interesting candidate for an \emph{efficient} quantum algorithm. In fact it is in NP but is not believed to be NP-Complete: like factoring, it belongs to the  NP-Intermediate family \cite{schoning88} and is representative of the (non-Abelian) \emph{hidden subgroup} problem family \cite{childs:qa,hallgren:hsp,moore:sfs_sym}. The best classical general algorithm solves GI for graphs of $n$ vertices in time $O(c^{\sqrt{n}\log n})$, were $c$ is a constant. 
\\[5pt]One way to solve GI is to show that two graphs are non-isomorphic. Starting from 2005 there have been different proposals of quantum algorithms based on  ``non-isomorphism witnesses'', i.e. observable quantities that assume different values only if the two input graphs are non-isomorphic. The standard benchmark for this approach is provided by the family of Strongly Regular Graphs (SRGs), that includes many hard instances of GI \cite{spielman96}. For example in  \cite{shiau05, gamble10, gamble12} to distinguish non isomorphic graphs the authors exploit continuous \cite{farhi03,childs02,keating06} and discrete time quantum walks \cite{Kempe03} of one or more particles moving through the graphs and compare the evolution of the same initial condition on the two graphs. The distinguishing power of the algorithm increases with the number of walker moving along the graph; the technique, however, is not universal and there are non-isomorphic graphs that cannot be distinguished.
\\A different approach, based on the Adiabatic Quantum Computation paradigm (AQC)\cite{farhi01,farhi00}, has been recently proposed in \cite{hen12,gaitan13}. In order to distinguish non-isomorphic graphs, for example, Vinci {\it et al.} look at the values assumed by a set non-isomorphism witnesses during the adiabatic evolution of the couple of graphs under investigation. They show that their technique is able to distinguish non-isomorphic SRGs up to instances of 29 vertices. On the other side, the technique is not guaranteed against the problem that afflicts all the quantum algorithm based on the adiabatic theorem: the spectral gap of the driving Hamiltonian can become exponentially small when the size of the problem increases; consequently, it could take an exponentially long time to reach the time-region in which it is possible to distinguish non-isomorphic graphs. 
Recently \cite{severini13} it has been shown that there is a family of observables that can be used to distinguish non-isomorphic graphs even if the ``adiabatic protocol'' is not respected and the systems  under observations are subjected to some degree of noise. An interesting feature of both Hen-Young  and Vinci {\it et al} proposals is that they can be, in principle, experimentally verified on current commercial hardware (D-Wave One \cite{rose11}). 
\\[5pt]In this work we propose an alternative approach to GI based on AQC. Instead of looking for non-isomorphism witnesses, the algorithm we propose solves GI by finding, if it exists, a permutation that transforms one of the two input graphs into the other. It uses a number of qubits that scales quadratically with the input size ($|V|=n$). The configuration space is explored through  a continuous-time quantum-walk  of $n$ interacting walkers that, by construction, visits only the space of \emph{functions} from $\{1,2,\ldots,n\}$ to $\{1,2,\ldots,n\}$. This makes it possible to define a cost function that is equivalent to a boolean formula made up of clauses of two literals ($2$-SAT), which can be easily turned into a $2$-local Hamiltonian, without using any perturbative gadget or projective reduction \cite{kempe06,jordan08}.
\\[5pt]The paper is organized as follows: in Section 2 we formally define the GI problem and the associated optimization problem. In Section 3 we  cast the optimization problem into an adiabatic algorithm. Section 4 is devoted to a presentation of the results. The last section is devoted to discussion, experimental verification proposal/issues and outlook. 
%
%
\section{Graph isomorphism as an optimization problem} \label{sec:cost}
An unoriented graph of size $n$ is a couple $G= (V,E)$, where $V=\{1,2, \ldots,n\}$ is set of vertices and two vertices $v,w \in V$ are connected to each other iff $\{v,w\}\in E$.
\\A permutation $\pi$ of the vertices is a bijection $\pi:V \to V$. We indicate by $\pi(G) = (V,E')$ the graph obtained by applying $\pi$ to $G$, where $E'= \{ \{\pi(v),\pi(w) \} : \{v,w\} \in E \}$. We will refer to the group of permutations of $n$ elements as to the \emph{symmetric group} $S_n$.
\\The \emph{Graph Isomorphism} problem (GI) is defined as follows: given two graphs $G_1 = (V,E_1)$ and $G_2=(V,E_2)$ of $n$ vertices, does exist a permutation $\pi \in S_n$ such that $G_2= \pi(G_1)$?
In what follows we will indicate the set of solutions of an assigned instance of GI as:
\[
 Iso(G_1,G_2) = \left\{ \pi \in S_n: G_2=\pi(G_1)\right \},
\]
and indicate $G_1 \cong G_2$ if $Iso(G_1,G_2)$ is non-empty.
\\We start our construction of a quantum algorithm for GI by defining a \emph{cost function} $f:S_n \to \mathbb{R}$ that assigns a penalty (positive weight) to every permutation not belonging to $Iso(G_1,G_2)$. Given the adjacency matrices $A_1$ and $A_2$ of, respectively, $G_1$ and $G_2$ and the permutation matrix $P_\pi$ associated to $\pi$, the function
\begin{align}
 f(\pi) = \frac{1}{2}\left | P_\pi A_1 P_\pi^T - A_2\right |_1,
\end{align}
counts the number of edges that are in $\pi(G_1)$ but not in $G_2$ and vice versa. Therefore $f(\pi)=0$ if $\pi \in Iso(G_1,G_2)$, $f(\pi)>0$ otherwise.
\\Instead of representing a permutation $\pi \in S_n$ through its permutation matrix $P_\pi$ we use a set of $n^2$  variables $\left (x_{1,1},x_{1,2},\ldots,x_{n,n} \right)$, $x_{i,j} \in \{ 0,1 \}, i,j=1,2,\ldots,n$, organized in a grid on $n$ rows and $n$ columns (see \fref{fig:system}). The variable $x_{i,j}$ is set to 1 if the permutation $\pi$ assigns to the element at position $i$ the element at position $j$.  
\\With this representation, the cost function $f$ becomes a real valued function $f:\{0,1\}^{n^2} \to \mathbb{Z}^+ \cup\{0\}$:

\begin{align} \label{eq:cost}
 &f \left (x_{1,1},\ldots,x_{n,n} \right) =   \\
& =\sum_
 {\scriptsize
 \begin{array}{c}
 \{i,j\} \in E_1\\
 \{k,l\} \notin E_2
 \end{array} 
 }
 x_{i,k} \ x_{j,l} +
  \sum_
 {\scriptsize
 \begin{array}{c}
 \{i,j\} \notin E_1\\
 \{k,l\} \in E_2
 \end{array} 
 }
x_{i,k} \  x_{j,l} + \nonumber \\
 & +\sum_{i=1}^n\left |1-\sum_{j=1}^n x_{i,j}\right |+ \sum_{j=1}^n \left |1-\sum_{i=1}^n x_{i,j}\right| \nonumber.
\end{align}
The addenda in the last line assign a penalty to every configuration that do not correspond to a permutation, i.e. has more than one 1 in each row and column.
\\Finding an assignment to the variables $x_{i,j}$ such that $f \left (x_{1,1},\ldots,x_{n,n} \right)=0$ is equivalent to the problem of finding a satisfying assignment to the following boolean CNF formula:
\begin{align}\label{eq:sat}
 &\bigwedge_{\substack{\{i,j\} \in E_1 \\ \{k,l\} \not\in E_2}} (\bar{x}_{i,k} \lor \bar{x}_{j,l}) \quad \wedge \bigwedge_{\substack{\{i,j\} \not\in E_1 \\ \{k,l\} \in E_2}} (\bar{x}_{i,k} \lor \bar{x}_{j,l}) \quad \\
 &\wedge \quad \bigwedge_{\substack{k \\i \neq j}} (\bar{x}_{i,k} \lor \bar{x}_{j,k}) \quad \wedge \quad \bigwedge_i \llrr{x_{i,1} \lor \ldots \lor x_{i,n}} \nonumber.
\end{align}
This is an $n$-SAT formula. The terms in the first row of \eref{eq:sat} are 2-literal clauses and depend on the input graphs; the terms of the second row are simultaneously satisfied only if there is exactly one ``1'' in each row and column of the grid: the $n$-literals terms $\llrr{x_{i,1} \lor \ldots \lor x_{i,n}}$ are satisfied as long as there is at least one ``1'' in each row, whereas the term $ \bigwedge_{\substack{k \\i \neq j}} (\bar{x}_{i,k} \lor \bar{x}_{j,k})$ is satisfied if there is at most one ``1'' in each column. To sum up, the second line of the formula \eref{eq:sat} is evaluated to \emph{true} is if the variables in the grid form a \emph{permutation matrix} and the first line is \emph{true} if such permutation maps one of the input graphs into the other, i.e. $G_1 \cong G_2$. We observe that, if we restrict the possible assignments to the variables $\{x_{i,j}\}_{i,j=1}^n$ to those corresponding to configurations in which there is exactly one ``1'' in each row of the grid, all the $n$-literal clauses will be automatically satisfied and the satisfaction of the formula $\bigwedge_{\substack{k \\i \neq j}} (\bar{x}_{i,k} \lor \bar{x}_{j,k})$ alone would guarantee that the configuration of the grid corresponds to a permutation. Under this assumption, the cost function $f$ is equivalent to the $2-SAT$ formula:
\begin{align}\label{eq:twosat}
 &\bigwedge_{\substack{\{i,j\} \in E_1 \\ \{k,l\} \not\in E_2}} (\bar{x}_{i,k} \lor \bar{x}_{j,l})
 \quad 
 \wedge \bigwedge_{\substack{\{i,j\} \not\in E_1 \\ \{k,l\} \in E_2}} (\bar{x}_{i,k} \lor \bar{x}_{j,l}) \quad \\
& \wedge \quad \bigwedge_{\substack{k \\i \neq j}} (\bar{x}_{i,k} \lor \bar{x}_{j,k}) \nonumber,
\end{align}
i.e. a formula made up of terms involving at most (in our case, exaclty) two variables. This fact will play a central role in the construction of the following section. 
\section{Adiabatic quantum walk} \label{sec:adiabatic}
The solution of a combinatorial problem, such as GI, can be mapped into the state of lowest energy of a potential operator, or \emph{final} Hamiltonian, $H_f$ \cite{apo89}.  In AQC the problem of finding such a state is solved by using an auxiliary, or \emph{initial} Hamiltonian $H_I$. The system is prepared in the ``easy to prepare'' ground state of $H_I$ and evolves under the action of a time-dependent Hamiltonian of the form:
\begin{align} \label{eq:adiabatic}
H(t)&= \llrr{1-\frac{t}{T}} H_I + \frac{t}{T} H_f, \qquad t \in [0,T].
\end{align}
If the evolution time $T$ satisfies
\begin{align} \label{eq:annealingtime}
 T  \gg \frac{\epsilon}{g_{min}^2}, 
\end{align}
with $\epsilon$ and the spectral gap $g_{min}$ defined as in \cite{farhi00} (see also Appendix A), the hypothesis of the adiabatic theorem are satisfied and the state of the system at the final time $T$ will be the ground state of $H_f$.
\\[5pt]In order to turn the optimization problem defined in the previous section into a quantum algorithm, we first assign a two-level system (qubit) \mbox{$\bm{\sigma}_{i,j} = \llrr{\sigma_{i,j}^x,\sigma_{i,j}^y,\sigma_{i,j}^z}$} to each boolean variable $x_{i,j}$ (see \fref{fig:system}). We select the direction $z$ as the computational direction and indicate by $\ket{+1}_{i,j}$ (or ``up'') and $\ket{-1}_{i,j}$ (or ``down'') the eigenstates of $\sigma_{i,j}^z$ belonging to the eigenvalues +1 and -1 respectively.
\\[5pt] The conventional generator of the diffusion ($H_I$ in \eref{eq:adiabatic}) adopted in AQC, adapted to our system, has the form 
\[
D=\sum_{i,j=1}^n \sigma_{i,j}^x. 
\]
The ground state of the Hamiltonian $D$ is easy to prepare (all the spins aligned along the $x$ axis) and corresponds to an uniform superposition of all the possible configurations $\left \{ \ket{b_{1,1},\ldots,b_{n,n}},b_{i,j} \in  \{-1,1\},i,j=1,\ldots,n \right \}$.
\\On the other side, we observed that, by restricting the set of possible assignments, GI can be mapped to a 2-SAT formula. Consider then the Hamiltonian
\begin{align} \label{eq:initialHam}
 H_I = -\frac{1}{2} \sum_{i=1}^n \sum_{j=1}^{n-1} \llrr{\sigma_{i,j+1}^+\sigma_{i,j}^- + \sigma_{i,j}^+\sigma_{i,j+1}^-},
\end{align}
where $\sigma^\pm$ are the spin raising and lowering operators $\llrr{\sigma^x \pm i \sigma^y}/2$. Each row of the spin grid evolves independently of the others and the interactions in each chain are next-neighbors of $XY$ type. The number of spins ``up'', or excitations, in each chain is preserved by $H_I$; in fact, defined the \emph{number} operator for each chain as:
\[
N_i^z = \sum_{j=1}^n \frac{1+\sigma_{i,j}^z}{2}, \qquad i=1,2,\ldots,n
\]
it is $[N_i^z,H_I]=[N_i^z H_I - H_I N_i^z]=0$. In particular, if we choose, for each row $i$, an initial condition in the $N_i^z=1$ sector of the Hilbert space, the evolution under $H_I$ will remain into the space $\mathcal{H}_{|N_i^3=1}$, i.e. the space of functions from $\{1,2,\ldots,n\}$ to $\{1,2,\ldots,n\}$. Indeed, this property will be preserved as long as the Hamiltonian of the system has the form $\alpha H_I + \beta V,\ \alpha,\beta \in \mathbb{R}$, with $V$ diagonal in the computational basis. Moreover, the ground state of $H_I$ is easy to prepare either by an adiabatic scheme (see Appendix B) or by dissipative means \cite{cormick13}.
%
%
%
\\The operator $H_I$ restricted to  $\mathcal{H}_{|N_i^3=1}$ can be rewritten as:
\begin{align}\label{eq:initialHamprime}
H_{I}= -\frac{1}{2} \sum_{i=1}^n \sum_{j=1}^{n-1} \llrr{\left | j+1 \right \rangle_i \left \langle j \right|_i + \left | j \right \rangle_i \left \langle j+1 \right|_i},
\end{align}
where $\left | j \right \rangle_i$ indicates that the excitation of the $i$-th chain is at position $j$. 
\\Thanks to this simplified notation, it becomes clear that the exploration of the space of configuration is performed through $n$ continuous time quantum walks on linear graphs.
\\[5pt]In this setting, it is possible to formulate the GI problems in terms as in \eref{eq:twosat}. The formula can be translated into the following  potential Hamiltonian:
%
%
\begin{figure} 
\includegraphics[width=0.9\columnwidth]{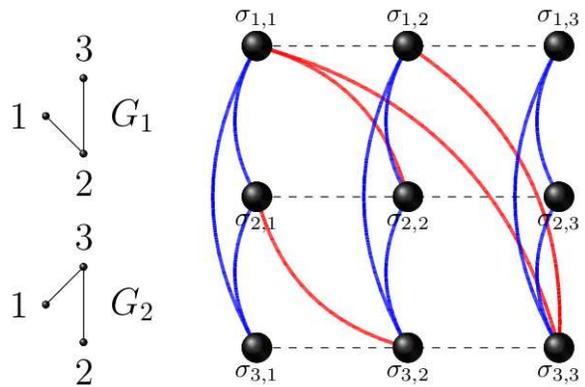}
\caption{(Color online) On the left, an instance of size $n=3$ of GI. On the right, the spin-grid and interaction graph for the same GI instance: solid lines correspond to $ZZ$ interactions: in blue we show the ``permutation''-constraints related interaction; in red, the instance dependent one. Dashed lines represent $XY$ interactions. \label{fig:system}}
\end{figure}
\begin{figure} 
\includegraphics[width=0.9\columnwidth]{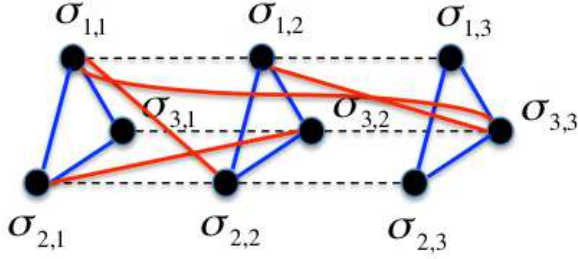}
\caption{(Color online) The same interaction-graph (and same color/style mapping) of \fref{fig:system} rearranged in order to show the topology of the \emph{hardware} part of the algorithm.\label{fig:systemgeom}}
\end{figure}
\begin{align} \label{eq:Hf}
 H_f &=\sum_
 {\scriptsize
 \begin{array}{c}
 \{i,j\} \in E_1\\
 \{k,l\} \notin E_2
 \end{array} 
 }
 \llrr{\frac{I+\sigma_{i,k}^z}{2}}  \llrr{\frac{I+\sigma_{j,l}^z}{2}}+  \\
&+  \sum_
 {\scriptsize
 \begin{array}{c}
 \{i,j\} \notin E_1\\
 \{k,l\} \in E_2
 \end{array} 
 }
 \llrr{\frac{I+\sigma_{i,k}^z}{2}}  \llrr{\frac{I+\sigma_{j,l}^z}{2}}+ \nonumber \\
 & +\sum_{i=1}^n \sum_{j=1}^n \sum_{k=j+1}^n \llrr{\frac{I+\sigma_{j,i}^z}{2}}  \llrr{\frac{I+\sigma_{k,i}^z}{2}}. \nonumber
\end{align}
The Hamiltonian is 2-local (i.e. it is made up of terms involving at most two qubit). To every violated clause in \eref{eq:twosat} it corresponds a unit energy penalty. If the 2-SAT formula associated to the GI instance $(G_1,G_2)$ is satisfiable, i.e. $G_1 \cong G_2$, there exists a zero energy configuration. 
\\The topology of the ensuing interaction graph has particular features. Within each chain there are only next-neighbor interactions. The $ZZ$ interactions  between the spins in a column of a grid, on the other side, define a complete $n$-graph. Together, the \emph{infra}-chain and \emph{infra}-column allow for the search of the solution to happen close to the space of permutations: they depend on the input size $n$ alone, and not on the particular instance of GI: they represent the \emph{hardware} part of the algorithm. The ``geometry'' that minimizes the ``distance'' of the hardware part is that of a cylinder. 
\\The instance-dependent interactions connect only elements that sit on different rows and columns: they  must be programmed ad-hoc (\emph{software}).  \Fref{fig:system} shows an example of the interaction-graph associated to a GI instance of dimension $n=3$.
\\If started from the ground state of $H_I$, restricted to $\mathcal{H}_{|N_i^3=1}$, the adiabatic evolution (i.e. with $T>\epsilon/g_{min}^2$) of the system under the action of the time-dependent Hamiltonian \eref{eq:adiabatic}, with $H_I$ and $H_f$ defined as in \eref{eq:initialHamprime} and \eref{eq:Hf}, will end up in the ground state \ket{e_0(T)} of $H_f$ (see Appendix A).  If $G_1$ and $G_2$ are non-isomorphic, the ground state energy will be equal to the number of clauses that cannot be satisfied, i.e. $\geq 1$. 
\section{Reading the output}
 We will address the key issue of the estimation of the ``annealing time'' $T$ in the next section. Here we propose a measurement protocol for the read-out of the result.
\\First of all, we observe that the expectation value $\bra{e_0(T)}C \ket{e_0(T)}$ of the observable
\begin{align}
C &=\sum_
 {\scriptsize
 \begin{array}{c}
 \{i,j\} \in E_1\\
 \{k,l\} \notin E_2
 \end{array} 
 }
 \llrr{\frac{I+\sigma_{i,k}^z}{2}}  \llrr{\frac{I+\sigma_{j,l}^z}{2}}+  \\
&+  \sum_
 {\scriptsize
 \begin{array}{c}
 \{i,j\} \notin E_1\\
 \{k,l\} \in E_2
 \end{array} 
 }
 \llrr{\frac{I+\sigma_{i,k}^z}{2}}  \llrr{\frac{I+\sigma_{j,l}^z}{2}}+ \nonumber 
\end{align}
is an isomorphism witness. In fact, it is zero iff the two graphs are isomorphic. 
\\Besides, the final state of the computation carries informations on $Iso(G_1,G_2)$, even in the case the observable $C$ cannot be measured. For example, if the input graphs are \emph{rigid}, i.e. the group of automorphisms of each of the graphs consists of the identity alone \cite{annot2}, then there is at most one solution to GI and $\|Iso(G_1,G_2)\| \leq 1$. The ground state of $H_f$, therefore, either encodes the permutation that maps $G_1$ into $G_2$ or not. By performing local and independent measurements of the \emph{position} observables 
\begin{align}
Q_i &= \sum_{x=1}^n \frac{1+\sigma_{i,x}^z}{2}, \quad i=1,2,\ldots,n,
\end{align}
we can read out the permutation $\ket{\pi} = \ket{\ves{q}{n}}$; then it suffices to check that  $\pi(G_1)=G_2$. 
\\If the graphs are not rigid, the ground state will be a superposition
\[
\sum_{i}\alpha_i \ket{\pi_i},\quad \pi_i \in Iso(G_1,G_2), \quad \sum_i |\alpha_i|^2=1.
\]
In order to extract one of the solution, we can proceed as follows. We run the algorithm once. We measure $Q_1$. The measurement will provide the value $q_1$. We then restart the algorithm by setting the ``spin up'' of the first chain to $q_1$, while the state of the other chains is prepared in the ground state of 
\[
H_{I}^{(1)}= -\frac{1}{2} \sum_{i=2}^n \sum_{j=1}^{n-1} \llrr{\left | j+1 \right \rangle_i \left \langle j \right|_i + \left | j \right \rangle_i \left \langle j+1 \right|_i}.
\]
We then let the system evolve under
\[
H^{{1}}(t)=\llrr{1-\frac{t}{T}} H_I^{(1)} + \frac{t}{T} H_f, \qquad t \in [0,T].
\]
We then measure $Q_2$ and iterate the procedure. After $n$ iterations of this scheme, we will end up in a permutation state $\ket{\pi} = \ket{\ves{q}{n}}$ and it suffices to  verify whether it maps $G_1$ into $G_2$ to have a definite answer.
So, in the case the input graphs are not guaranteed to be rigid, we need at most a linear time overhead in order to read out the result and the overall execution time of the algorithm (adiabatic procedure + measurement) will scale, in the worst case, as $O(nT)$, $T$ being the annealing time required by the first run of the adiabatic procedure.
\\Independently of their rigidity of the input graphs the output of the algorithm is always a permutation $\pi$. If the two input graphs are not isomorphic, it will be $G_1 \neq \pi(G_2)$. In what follows, therefore, we will restrict our investigation on the performance of the algorithm on isomorphic instances of GI.
\section{Results}
\begin{figure}[t]
\begin{center}
\subfigure[]{\label{fig:figura3a} \includegraphics[width=0.8\columnwidth]{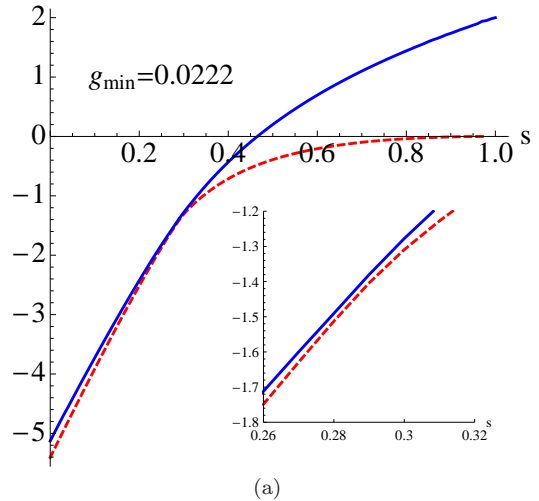}}
\subfigure[]{\label{fig:figura3b} \includegraphics[width=0.8 \columnwidth]{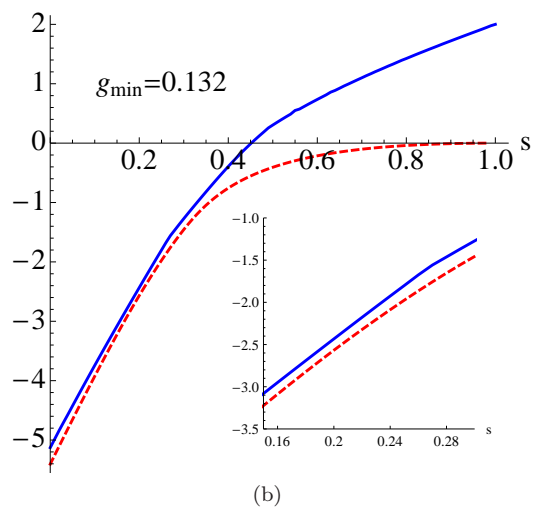}}
\caption{A case study. Graph size $n=6$. (a) The ground state energy (dashed line) and the first excited energy (solid line) for a random instance $(G_1,\pi(G_1))$. Inset: an expanded view of the critical region: there is no level crossing. (b) The same quantities for an instance $(G_1, \sigma(\pi(G_1)))$. The spectral gap is one order of magnitude larger than for the original instance $(G_1, \pi(G_1)$. Inset: an expanded view of the critical region.}
\label{fig:AQC}
\end{center}
\end{figure}

\begin{figure}[]
\begin{center}
\includegraphics[width=\columnwidth]{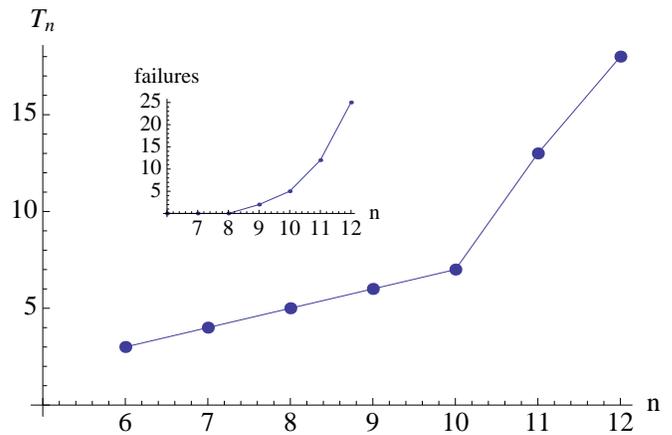}
\caption{The annealing time $T_n$ as a function of the problem size $|V|=n$. In the inset: the number of failures on 100 instances, for an annealing time determined as in the main figure, without the permutation trick. }
\label{fig:QMC}
\end{center}
\end{figure}

\begin{figure}[]
\begin{center}
\includegraphics[width=\columnwidth]{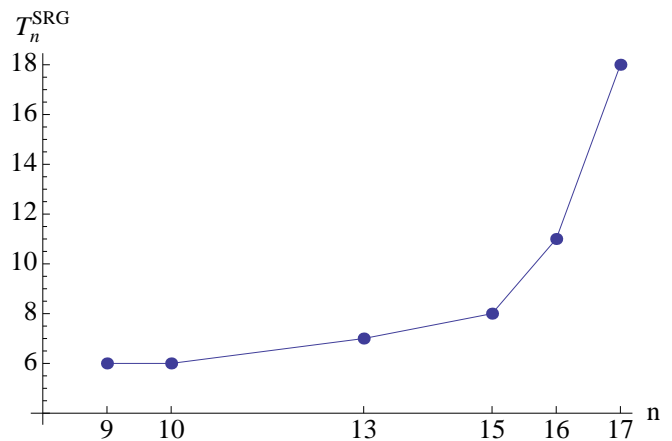}
\caption{The annealing time $T_n^{SRG}$ for strongly regular graphs as a function of the problem size $|V|=n$. }
\label{fig:QMCSRG}
\end{center}
\end{figure}

For $t>0$, the spin chains interact with each other. The analysis of the spectral gap of the Hamiltonian \eref{eq:adiabatic} is quite hard. We did not find any mean to derive analytic results about the spectral gap $g_{min}$; we can only warrant that it $g_{min}>0,\qquad t \in [0,T]$.
The result follows immediately by an application of the Perron-Frobenius theorem \cite{johnson12}.
\\This, together with the results about the spectra of the operators $H_I$ and  $H_f$ of the previous sections,  assure that the algorithm ``makes sense'' but provides no information about its efficiency: we cannot rule out the possibility of $g_{min}$ becoming exponentially small as the input size increases.
\\The determination of the spectral gap for instances of size $n$ requires the  solution of the eigenvalue problem for $n^n \times n^n$ matrices by numeric means. With our computational resources, we have been able to characterize the spectral gap for  graphs of at most $n=7$ vertices (i.e., for a system of $49$ qubits, evolving in a Hilbert space isomorphic to $\mathbb{C}^{823543}$). This does not allow for a study of the spectral behavior of the algorithm as a function of the input size \cite{krovi10,karimi11}. By direct inspection  of the spectral gap, however, it is easy to see that the ``hardness'' (i.e. $g_{min}$) of an isomorphic instance $(G,\pi(G)),\ \pi \in S_n$, of GI may depend on $\pi$ (see \fref{fig:AQC}).
\\The observation of this simple ``fact of life'' suggests the following strategy, that we christened \emph{Permutation Trick} (PT): try to solve the original instance $(G,\pi(G))$. If at the end of the adiabatic evolution a solution is not found, modify the input instance $(G,\pi(G)) \to (G,\sigma(\pi(G)))$, with $\sigma \in S_n$.
\\[5pt]For more significant instances we resorted to Monte-Carlo simulations. We used  the World-Line Quantum Monte-Carlo (QMC) \cite{troyer03} numerical scheme. The algorithm is described in the Appendix C.
%
\\In order to study the dependence of the annealing time $T_n$ on the problem size $n$ we proceeded as follows. We generated a sample of $N=100$ isomorphic instances $(G_1^i,G_2^i=\pi_i(G_1^i)),i=1,2,\ldots,N$, with $\pi_i$ randomly extracted from the symmetric group $S_n$; each of the graphs $G_1^i$ is connected and is generated using the Wolfram Mathematica function {\bf RandomGraph($\mathbf{\{n,m_i\}}$)} (and discarding  non-connected graphs); the parameter $m_i$ is the number of edges of the graph, uniformly extracted in the range $[2n,n(n-2)/2]\cap \mathbb{Z}$: we avoided graphs with low connectivity, since they usually provide very easy GI instances.
\\For each instance we ran the QMC simulation for a tentative time, say $T'$, and up to $h=5$ times. If a solution is found, stop;  otherwise, apply the Permutation Trick: sample $\sigma \in S_n$ and try to solve $(G,\sigma(\pi(G)))$. The algorithm fails when a solution in not found after $k=4$ applications of the permutation trick. We point out that the maximum number of Monte-Carlo runs for each instance is $k \cdot h=20$ independently of the instance size. We define the ``annealing time'' $T_n$ as the time needed to solve \emph{all} the $N$ instances of size $n$ of GI, with the help of PT.  The results are shown in \fref{fig:QMC}. We show also the number of failures of the algorithm when we run it on instances of size $n$ with annealing time $T_n$,  $h=50$ and without the application of PT ($k=0$); the steep growth of the number of failures supports the conjecture that a rearrangement of the ``solution landscape'' is likely to significantly simplify the original instance, without modifying its structural properties.
\\In order to avoid any misunderstanding, we stress here that the results we will discuss below are inconclusive under, at least, two points of view. First, the dimension of the instances is very limited. Secondly, the QMC simulation of the adiabatic scheme is not guaranteed to provide a faithful simulation of the evolution of the system \cite{hastings13}. What we are presenting here, therefore, are preliminary results and observations.
\\The results obtained with QMC for random graphs of size up to $n=12$ vertices are shown in \fref{fig:QMC}. The annealing time scales linearly from $n=6$ to $n=10$. Then there is some kind of ``phase transition'': the time required to solve instances of size $n=11$ is about twice the time needed to solve the $n=10$ instances. Then the annealing time grows linearly from $n=11$ to $n=12$ (but more steeply than from $n=6 \to 10$).  
\\Needless to say, the reduced size of the tractable instances makes it impossible to infer anything about the behavior of the algorithm on large GI instances. The presence of ``phase transitions'', like the one observed at $n=10\to11$, will most likely imply an exponential dependence of the annealing time on the input size; the rate of such transitions, however, will determine the presence of any quantum speed-up with respect to the best classical algorithm.
\\Since SRGs can provide harder instances of GI \cite{spielman96}, we tested our algorithm on instances of GI generated from SRG up to $n=17$ vertices. The class of SRG is organized in families $(n,k,\lambda,\mu)$, where $n$ is the number of vertices, $k$ is the degree of each vertex, $\lambda$ is the number of common neighbors of any two adjacent vertices and $\mu$  is the number of common neighbors shared by any two non-adjacent vertices. Unfortunately the families $(n,k,\lambda,\mu)$  are made up of at most two representatives for $n \leq 17$. In order to allow for the comparison with the results obtained with randomly generated graphs to be fair, for each representative $G$ of the SRG family $(n,k,\lambda,\mu)$, we generated 10 instances $(G,\pi_i(G))$, with $\pi_i$ randomly extracted from $S_n$; as for random graphs, we define $T_n^{SRG}$ as the time needed to solve all the 10 $n$-instances. The results are shown in  \fref{fig:QMCSRG}. The annealing times for SRG are usually much smaller than those needed to solve random graphs. This allowed us to push the QMC simulations with SRG up to instances of $n=17$ vertices. A direct comparison with the annealing-time required by random graphs is possible for $n \in \{9, 10,13\}$. For $n \notin \{9, 10,13\}$ we found instances of random graphs of size $n$ that are not solved for $T=T_n^{SRG}$, thus showing that $T_n>T_n^{SRG}$. As far as small graphs are considered, therefore, strong regularity is an advantage.

%

\section{Conclusion, outlook and Experimental verification}
The $2$-local quantum adiabatic algorithm for GI we presented finds the isomorphism between the input graphs by finding, if it exists, a permutation matrix that maps one of the two graphs into the other. By using $n$ interacting quantum walks, we were able to reduce the GI problem to the search of a satisfying assignment to a $2$-SAT formula. Remarkably enough, this is done without resorting to any perturbation gadget or projective technique. 
\\The algorithm is a true quantum algorithm. In fact, the initial Hamiltonian  $H_I$ (actually a slightly modified version of it, see Appendix B) is frustration-free and stoquastic \cite{bravyi08}. When the two input graphs $G_1,G_2$ are isomorphic, and it is the case for all the instances used in our study, the final Hamiltonian $H_F$ is frustration-free and stoquastic as well. On the other side, while $H(t),\ 0 < t < T$, preserves the stoquasticity, it is no guaranteed to be frustration-free. This rules out the possibility of efficiently simulate our algorithm by classical means \cite{bravyi09}.
\\[5pt]We cannot provide a characterization of the spectral behavior \cite{laumann12} of the adiabatic Hamiltonian driving the system; in the lack of analytic results, we resorted to numerics, which allow for an inspection of the spectral gap only for GI instances up to $n=6$ vertices, which is obviously largely insufficient to infer any scaling law. 
\\With the help of Monte-Carlo simulations we were able to get some preliminary results about the running-time of the algorithm for random  graphs and SRG. There is no evidence of any quantum speed-up with respect to the best classical algorithm for GI. In fact, the (admittedly very limited) data on the annealing-times $T_n$ and $T_n^{SRG}$, needed to solve, respectively, random and strongly regular graphs, fit very well to a scaling $O(2^{\sqrt{n}\log(n)})$. If the scaling were confirmed by an extended simulation campaign, we could therefore only claim (no surprise here) that the adiabatic procedure we defined is not equivalent to a Grover search \cite{grover96} in the \emph{unstructured} $n^n$- dimensional space of functions from $\{1,2,\ldots,n\}$ to $\{1,2,\ldots,n\}$, nor in the  $n!$-dimensional space of permutations, since it would require a time $O(2^{\frac{n}{2}\log(n)})$. 
\\From the point of view of complexity, therefore, the results we obtain are quite modest, but maybe not unexpected. AQC offers the potential advantage of being  a general purpose tool; as such it may be not the \emph{best} tool for any given problem. The 2-SAT problem, to which we reduce GI in our setting, provides a key example: it is in the complexity class $P$, since there is an \emph{ad hoc} algorithm that solves it in linear time. A 2-SAT problem can be straightforwardly encoded into a 2-local Hamiltonian by a construction similar to the one presented in Section II and be used as the final Hamiltonian of an Adiabatic Algorithm. In the AQC setting, however, the adiabatic Hamiltonian to solve 2-SAT is equal to the one used to solve NP-Hard problem MAX-2-SAT, that is the problem of determining the maximum number of 2-literal clauses that can be simultaneously satisfied \cite{garey76}.
  It is possible that satisfiable 2-SAT formulas or, equivalently, isomorphic instances $G_1 \cong G_2$, are easier to solve than unsatisfiable (non-isomorphic) instances: in this case, in fact, the final Hamiltonian is frustration-free. This conjecture, however, remains to be proved. 
\\[5pt]
The results we obtained through Monte-Carlo simulations must be considered with caution: it is possible that the numerical scheme (and the parametrization) we used does not capture some fundamental aspect of the quantum adiabatic evolution. Besides, the simulations must be pushed much further to understand, in the spirit of \cite{karimi11}, the real dependence of the annealing time on the size of the instances.  The development of optimized and parallelized quantum Monte-Carlo algorithms, exploiting the computational power of multi-core CPU and GPUs, will be one of the focuses of future research.  However, the dimension ${n^n}$ of the Hilbert space visited by our algorithm is such that, even by exploiting all the computational resources used in Ref.\cite{karimi10}, we will be able to simulate the algorithm for graphs of at most $n \approx 25$ vertices.  A real check of the performance of the procedure described in this work will be possible only by implementing it to a quantum computational device.
%
%
\\[5pt]We thus conclude with a discussion of the difficulties one would encounter in an hardware implementation of the algorithm.
\\As a matter of example, let us consider the D-Wave One quantum computer \cite{rose11}.  The fact that the device implements the standard AQC paradigm, and promises to be easily scalable, makes it look as an ideal candidate for an experimental verification of our procedure. 
\\The main issue with this reference architecture is related to the kind of interactions required by the algorithm. The current version of the device does not implement $XY$-interactions. As a matter of fact the D-Wave One is currently able to solve only problems that can be mapped into a 2-D Ising problem, that is problems that can be mapped to standard AQC Hamiltonians involving only $ZZ$ interactions between nearest-neighbor qubits and a transverse field $\sigma^x$. On the other side the superconducting flux-flux qubits used in the D-Wave One can in principle support $XY$ interactions \cite{chancellor13}, so it is possible that our scheme will become implementable in some next-generation version of the hardware.
\\Another, somehow minor, criticality is the mapping of the interaction-graph  determined by the algorithm (see figures \ref{fig:system} and \ref{fig:systemgeom}) onto the Chimera-Graph (see, for example, figure 1 of Ref.\cite{perdomo12} for a representation of the graph). The \emph{Minor-Embedding} procedure \cite{choi11} can map a complete graph onto the Chimera Graph with a quadratic resource overhead. This means that our interaction graph can be mapped into the D-Wave graph; what remains to be understood is the effect that such an embedding will induce on the execution time of the algorithm.
\\In other physical implementations, such as crystal of trapped ions \cite{monroe02}, the realization of the $XY$-Hamiltonian, together with its control and the preparation of its ground state, will be quite straightforward. In this setup, however, it is the realization of the $ZZ$ interactions between distant qubits that may be very challenging, and would require some sort of \emph{quantum bus} \cite{brennen03}. The definition verification scheme for our algorithm based on current technology and will be the focus of future research.


\begin{thebibliography}{42}
\bibitem [{ann()}]{annot2}%
  \BibitemOpen
  \href@noop {} {\enquote {\bibinfo {title} {The problem of determining the
  group of automorphisms of a graph is (almost) as difficult as gi.}}\
  }\BibitemShut {NoStop}%
\end{thebibliography}
%
%
%
%
\section*{Appendix A}
For the sake of self-containedness, we report here some basic definitions related to the adiabatic theorem.
\\Given two Hamiltonian operators $K$ and $V$ on  on $\mathbb{C}^{n}$, let us consider the time-dependent Hamiltonian
\begin{align}
 H(s) = (1-s) K + s V, \qquad 0 \leq s \leq 1.
\end{align}
We indicate by $e_0(t) < e_1(t) < \ldots < e_n(s)$ and $\ket{e_0(s)},\ket{e_1(s)},\ldots,\ket{e_n(t)}$ the instantaneous non-degenerate eigenvalues of $H(s)$ and the corresponding eigenvectors.
\\The \emph{spectral gap} of $H(s)$ is defined as
\[
 g_{min}= \min_{0 \leq s \leq 1}\llrr{e(1)-e(0)}.
\]
The adiabatic theorem asserts that, if the rescaling constant $T$ satisfies the relation $T \gg \epsilon/g_{min}^2$, where
\[
\epsilon = \max_{0 \leq s \leq 1} \left | \bra{e_1} \frac{d H(s)}{ds} \ket{e(0)} \right |,
\]
then a system prepared at time $t=0$ in the ground state of $H(0)=K$ will follow the instantaneous ground state $\ket{e(t)}$ of the rescaled Hamiltonian $H(s \cdot T)$ and end up, at time $t=T$ in the ground state of the Hamiltonian $V$.
\\While the value $\epsilon$ can be usually bounded from above by a polynomial in the system size $n$, the spectral gap $g_{min}$ can happen to have an exponential dependence on the system size.
\section*{Appendix B}
The preparation of the ground state of the initial Hamiltonian $H_I$ (see equation \ref{eq:initialHam}) restricted to the $N_3=1$ sector of the Hilbert space of the system can be done efficiently by adiabatic means.
\\In what follows we will describe the preparation of a single chain of the system. The overall initial state will then be obtained by tensorialization.
\\Consider the initial state
\[
\ket{\sigma_{1}^3 = -1,\ldots,\sigma_{\lfloor \frac{n+1}{2}\rfloor}^3 =+1,\ldots,\sigma_{n}^3 = -1},
\]
describing a chain with a single spin up at position $\lfloor \frac{n+1}{2}\rfloor $. This is the ground state of the Hamiltonian
\begin{align} \label{eq:auxHam}
H_I^{aux} = -\frac{V}{2} \llrr{1+\sigma_{\lfloor \frac{n+1}{2}\rfloor)}^3}
\end{align}
for any $V>0$.
\\We let the system evolve under
\begin{equation}\label{eq:auxHam}
H^{aux}(t) = \frac{t}{T} H_F^{aux} + (1-\frac{t}{T}) H_I^{aux},
\end{equation}
where
\begin{align} \label{eq:hamXY}
H_F^{aux} &=  -\frac{1}{2} \sum_{j=1}^{n-1} H_F^{aux}(j,j+1) = -\frac{1}{2} \sigma_{j+1}^+ \sigma_{j}^- + \sigma_{j}^+ \sigma_{j+1}^-.
\end{align}
The annealing time depends polynomially on the system size $n$. In fact the spectral gap of \eref{eq:auxHam} can be analytically determined by standard techniques \cite{feynLec3} to be, for  \mbox{$0\leq s < 1$}
\[ 
\cos\llrr{\frac{2\pi}{n+1}}-\sqrt{\cos\llrr{\frac{\pi}{n+1}}^2 s^2 +\llrr{\llrr{1-s}V}^2}.
\]
The gap is monotonically decreasing in $s$ and reaches its minimum at $s=1$. 
For $s=1$ the gap is  the gap of the isotropic $XY$ on $n$ sites Hamiltonian restricted to the single excitation subspace $\mathcal{H}_{|N_3=1}$, that is
\[
\cos\llrr{\frac{2 \pi}{n+1}} - \cos\llrr{\frac{\pi}{n+1}} = \Omega \llrr{\frac{1}{n^2}}.
\]
The ground state of $H_F^{aux}$ can therefore be prepared efficiently.
\\We point out that while  \eref{eq:hamXY} is not frustration-free, it becomes such as soon as we add two localized potential. In fact the ground state of
\begin{align} \label{hamXYFF}
H_F^{aux,FF} &=  -\frac{1}{2} \sum_{j=1}^{n-1} H_F^{aux}(j,j+1) -\frac{1}{2}(\sigma_1^z+\sigma_n^z).
\end{align}
is the $W$ state 
\[
\ket{W}=\frac{1}{\sqrt{n}} \sum_{i=1}^n\ket{i}
\]
which minimizes $H_F^{aux}(j,j+1)$ for $j=1,2,\ldots,n-1$ and $-\sigma_1^z,\sigma_n^z$. 
\section*{Appendix C}
We use the  World-Line Quantum Monte-Carlo algorithm to simulate the evolution of the ground-state distribution of the $\sigma_{i,j}^z,\ i,j=1,\ldots,n$ observables. For a complete account on the numerical scheme, we refer the reader to \cite{scalettar98}. The $C$ code used to simulate the system is available at {\small \url{https://bitbucket.org/luca_zanetti/qmc_gi/downloads}}. Here we briefly describe the algorithm and define the parameters used in our simulations.
\\[5pt]We first discretize the time evolution. Instead of interpolating between $H_I$ \eref{eq:initialHamprime} and $H_f$ \eref{eq:Hf} by continuously varying the parameter $t$ (see \eref{eq:adiabatic}), we take an integer \emph{evolution time} $T$ and change the time-dependent system Hamiltonian through unit steps from 0 to $T$.
\\We approximate the evolution of the instantaneous ground state $\ket{e(k)} \to\ket{e(k+1)}$ of the system between two interpolation steps $k$ and $k+1$ via the Suzuki-Trotter replica method: $r$ replicas of the system are evolved through $m$ Metropolis moves toward the equilibrium distribution of $H(k+1)$ at temperature $1/\beta$. In our experimental campaign that the best results are obtained if we set $\beta=r$. 
\\The algorithm can be synthesized as follows:
\\[2pt] \

\begin{algorithmic}[0]
\State Read $G_1,G_2,T$
\For {$i=1, \dots, h$}
	\If {$i==1$}
		 {$\sigma_1 \gets 1,\sigma_2 \gets 1$}
	\Else
	 	{\ $\sigma_1,\sigma_2 \gets S_n \text{(uniformly at random)}$}
	\EndIf
	\State Initialize the $r$ replicas of each chain to the same configuration
	
	\For {$j \gets 1,\dots,k$}
		\State Thermal-annealing to  $e^{-\beta H_I}$
		\For {$t=1,\ldots,T$}
			\State Set $\nu=(1-\frac{t}{T})$
			\For {$l=1,\ldots,m$}
			\State  Metropolis Move with Hamiltonian $\nu H_I + (1-\nu) H_f$.
		\EndFor
		\EndFor
		\If {A fraction $\geq 1/6$ of the replicas has reached zero cost configuration}
			\State \textbf{return} Yes
		\EndIf
	\EndFor
\EndFor
\State \textbf{return} No
\end{algorithmic}
The thermal-annealing procedure is used to reproduce the equilibrium distribution of $e^{-\beta H_I}$ of the Hamiltonian \eref{eq:initialHamprime}.
The iterations over $i=1,\ldots,h$ implement the Permutation Trick. The iterations over $j=1,\ldots,k$ capture the non-deterministic nature of MC. 
\\Since $\beta<+\infty$ the thermal state will have a support larger than the sole ground state. Besides, the allowed number of Metropolis moves does not guarantee that the replicas equilibrate \cite{hastings13}. For these reasons, we say that the QMC procedure succeeds in finding a solution of an instance of GI when, at the final time $T$, $1/6$ of the replicas are in a configuration corresponding to a solution of the given instance. In this way we are able to capture the approximate nature of the solutions provided by the QMC numerical scheme, while ruling out the possibility of finding a solution by mere chance.
\\In our simulations the parameters have been set to: $h=5,\ k=4,\ r=200,\ m=250,\ \beta=m$.

\end{document}